\documentclass[journal]{IEEEtran}
\usepackage{cite}

\ifCLASSINFOpdf
   \usepackage[pdftex]{graphicx}
   \graphicspath{{../pdf/}{../jpeg/}}
   \DeclareGraphicsExtensions{.pdf,.jpeg,.png}
\else
   \usepackage[dvips]{graphicx}
   \graphicspath{{../eps/}}
   \DeclareGraphicsExtensions{.eps}
\fi
\usepackage{enumerate}
\usepackage{amsfonts}
\usepackage{booktabs}
\usepackage[cmex10]{amsmath}

\usepackage{mathabx}
\begin{document}
%
\title{Joint User Association and Power Control for Load Balancing in Downlink Heterogeneous Cellular Networks}
%
%
%

\author{\IEEEauthorblockN{Tianqing Zhou}\\
\IEEEauthorblockA{School of Information Science and Engineering, Southeast University, Nanjing 210096, China\\
Email: {zhoutian930}@163.com\\
}
}

\maketitle

\begin{abstract}
Instead of achievable rate in the conventional association, we utilize the effective rate to design two association schemes for load balancing in heterogeneous cellular networks (HCNs), which are both formulated as such problems with maximizing the sum of effective rates. In these two schemes, the one just considers user association, but the other introduces power control to mitigate interference and reduce energy consumption while performing user association. Since the effective rate is closely related to the load of some BS and the achievable rate of some user, it can be used as a key factor of association schemes for load balancing in HCNs. To solve the association problem without power control, we design a one-layer iterative algorithm, which converts the sum-of-ratio form of original optimization problem into a parameterized polynomial form. By combining this algorithm with power control algorithm, we propose a two-layer iterative algorithm for the association problem with power control. Specially, the outer layer performs user association using the algorithm of problem without power control, and the inner layer updates the transmit power of each BS using a power update function (PUF). At last, we give some convergence and complexity analyses for the proposed algorithms. As shown in simulation results, the proposed schemes have superior performance than the conventional association, and the scheme with joint user association and power control achieves a higher load balancing gain and energy efficiency than conventional scheme and other offloading scheme.
\end{abstract}

\begin{IEEEkeywords}
Heterogeneous cellular networks, user association, load balancing, power control, energy efficiency.
\end{IEEEkeywords}

%
\IEEEpeerreviewmaketitle

\section{Introduction}
%
%
%
%
\IEEEPARstart{W}{i}th the explosive growth of data traffic driven by various applications such as smartphones and tablets, the network operators have to find a good way of improving network capacity. The conventional macrocellular network is primarily designed to guarantee basic coverage and is clearly not a good solution to cope with this challenge \cite{1,2,3,4}. As the most promising solution to handle the data deluge, increasing the density of base stations (BSs) can reduce the frequency reuse distance and thus improve system capacity \cite{5}. Heterogeneous cellular networks (HCNs)--consisting of conventional macro BSs and heterogeneous elements such as pico BSs, femto BSs, distributed antennas and so on--have recently emerged as a cost-effective solution to handle the exploding and uneven data traffic demands.
\par
In HCNs, various BSs have significantly different transmit powers. When the conventional signal strength-based association scheme (e.g. maximal achievable rate association) is applied to HCNs, the obtained load distribution may be very imbalanced because most users are associated with high-power BSs and very few users can be attracted by low-power BSs. Such imbalance will mean that the resources of underloaded BSs cannot be fully utilized due to the limited number of associated users and some users may not be served by overloaded BSs because of insufficient resources. In order to improve the system performance of HCNs, i.e., fully utilize system resources and improve user experience, a user association scheme with offloading capability should be designed, which is also named as an offloading scheme. In addition, the users associated with low-power BSs often receive the strong interference from high-power BSs, which greatly degrades the system performance. To further improve user experience and reduce power consumption, joint user association and power control for load balancing in downlink HCNs should be a good option.

\subsection{Related work}
Unlike conventional cellular network, the design of effective association schemes for load balancing in HCNs becomes more difficult because different BSs coexist, and attracts more and more attention\cite{6,7}. In order to balance loads among different BSs, many kinds of offloading schemes are designed for HCNs.
\par
As a frequently utilized method, the biasing method \cite{8,9,10,11} balances the network loads by adding a biasing factor/offset to low-power BSs, where the factor/offset can theoretically narrow the power gap between high-power and low-power BSs. Although the biasing method is simple, it may be undesirable in the practical implementation because the optimal factor/offset with a closed-form expression cannot be found.
\par
To avoid searching the factor/offset, some efforts have been made to design other effective association strategies. In works \cite{12,13,14,15}, authors consider maximal sum-utility associations that can offload low-rate users from high-power BSs to low-power BSs, where the utility is denoted as a logarithmic function of long-term rates. Authors in works \cite{12,13,14} investigate the load balancing problem under equal bandwidth allocation, but authors in work \cite{15} study the same problem under co-channel deployment, orthogonal deployment and partially shared deployment. Unlike other works, authors in works \cite{16} and \cite{17} consider a minimal pathloss association and a repulsive cell activation respectively. The former is simple but offloads users in a relatively random manner, whereas the latter is complicated because the optimal minimal separation distance cannot be given in a closed form.
\par
In order to mitigate the across-tier and inter-tier interferences, some designers jointly consider user association and power control for uplink HCNs \cite{18,19}. In these association schemes, authors often minimize the uplink power consumption (the sum of uplink transmit powers) \cite{18} or the one mixed with other objective \cite{19}. Compared with uplink HCNs, the related works are fewer in downlink HCNs. In order to control the transmit powers of BSs and thus enhance system performance, many authors \cite{19,20,21} introduce the beamforming during user association. However, these association schemes may be unreasonable since the user association often takes place at a fairly long time scale but the beamforming takes place at a shorter time scale \cite{14}. Evidently, the user association utilizes a slow-fading channel, but the beamforming adopts a fast-fading channel. So far, the research on joint user association and (direct) power control may be just found in work \cite{14}, which optimizes a network utility maximization problem with power control.
\par
In addition, some authors \cite{22} are also in favour of power allocation during user association for downlink HCNs. Significantly, the power allocation often refers to that the transmit power of some BS is allocated to its associated users under QoS (quality of service) constraints, but the power control can be regarded as the change (decrement) of transmit power of some BS. Unlike the former, the latter doesn't often involve the QoS constraints. Evidently, the power allocation in the association problem may not guarantee the user fairness and waste resources. Specially, some users may not be selected (associated with BSs) when the QoS constraints cannot be met, and the allocated powers may be wasted when the associated users are not scheduled. Thus, the power allocation may be not a good option for the association problem.
\par
So far, there are few works that jointly consider user association and power control for downlink HCNs, and few works consider the user association that maximizes the sum of effective rates for downlink HCNs, and no effort jointly considers user association and power control to maximize the sum of effective rates for downlink HCNs. Considering that the effective rate is closely related to the the load (number of associated users) of some BS and the achievable rate of some user, it can be a key factor for load balancing, which is shown in works \cite{12,13,14,15}. Unlike the works \cite{12,13,14,15}, we consider the an association problem that maximizes the sum of effective rates but network utility. That is to say, we have a different perspective from the works \cite{12,13,14,15}, i.e., throughput maximization.
\subsection{Contributions and organization}
In this paper, we design two association schemes that maximize the sum of effective rates, and develop two effective association algorithms for the formulated problems. Specially, we make the following contributions in this paper.
\begin{enumerate}[1)]
\item \emph{User Association with Maximizing the Sum of Effective Rates (UAMSER):} We design an association scheme that maximizes the sum of effective rates, which is hardly considered in existing works. According to the form of the optimization problem, we can develop an effective one-layer iterative algorithm to achieve its solution, which converts the sum-of-ratio form of original optimization problem into a parameterized polynomial form.
\item \emph{Joint User Association and Power Control with Maximizing the Sum of Effective Rates (JUAPCMSER):} To further reduce energy consumption and mitigate interference, we introduce power control into the scheme UAMSER and thus obtain the scheme JUAPCMSER. However, the novel optimization problem is in a more complicated form. To solve it, we try to design an effective two-layer iterative algorithm that alternatively optimizes transmit powers of BSs and association indices. Specially, the outer layer performs user association using the one-layer iterative algorithm of scheme UAMSER, and the inner layer updates the transmit power of each BS using a power update function (puf).
\item \emph{Giving the Convergence and Complexity Analyses for Proposed Algorithms:} As for the proposed algorithms, we give the proofs of convergence for them, especially for association algorithm and power control algorithm. In addition, we also give some complexity analyses for proposed algorithms.
\end{enumerate}
\par
The rest of this paper is organized as follows. In Section 2, we give the system model, i.e., two-tier HCNs. In Section 3, we propose two association schemes including UAMSER and JUAPCMSER. In Section 4, we perform the numerical simulation to show the effectiveness of proposed schemes, and investigate some association performances such as load balancing gain, system throughput and energy efficiency. In Section 5, the conclusions are drawn.
\par
\noindent
\emph{Notations:} We denote $\boldsymbol{a}\succcurlyeq \boldsymbol{b}$ if ${{a}_{i}}\ge {{b}_{i}}$ for any $i$, and denote $\boldsymbol{a}\preccurlyeq \boldsymbol{b}$ if ${{a}_{i}}\le {{b}_{i}}$ for any $i$. In addition, we have $\left[ z \right]_{a}^{b}=\min \left\{ \max \left\{ z,a \right\},b \right\}$ and $\left[ z \right]^{+}=\max \left\{ z,0 \right\}$, and let $\log\left(z\right)$ be the logarithmic function with base $e\approx 2.7183$ in terms of parameter $z$.
\section{System model}
At present, there exist two types of deployments for HCNs, i.e., regular and irregular deployments \cite{7}. In the regular deployment, the MBSs are deployed according to a conventional cellular framework, but users and other low-power BSs are randomly scattered into each macrocell. Unlike the regular deployment, all users and BSs in the irregular deployment are deployed in a random manner. In addition, many works consider the Poisson point process (PPP) for modeling HCNs. In fact, it refers to the numbers of users and BSs obey the Poisson distribution. Whatever the deployment used for modeling HCNs, the essence of association scheme is unaffected by it. In other words, the performance differences among different association schemes are always consistent for different deployments.
\par
\begin{figure}[!t]
\centering
\centerline{\includegraphics[width=3.8in]{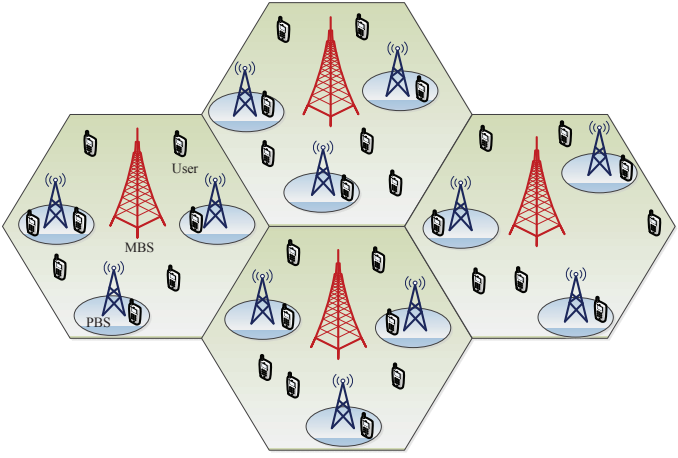}}
\caption{Illustration of two-tier HCNs consisting of PBSs and MBSs.}
\label{fig1}
\end{figure}

In this paper, we consider a two-tier HCN consisting of PBSs and MBSs and adopt the regular deployment for modeling it, which can be found in Fig. \ref{fig1}. Specially, the MBSs (macro BSs) are deployed into a cellular framework, while all users and PBSs are scattered into each macrocell in a random manner.
\par
We denote the set of BSs including MBSs and PBSs by $\mathcal{N}$, and represent the set of users as $\mathcal{K}$. In addition, we let cardinalities of sets $\mathcal{N}$ and $\mathcal{K}$ be $\left| \mathcal{N} \right|$ and $\left| \mathcal{K} \right|$ respectively. Then, the signal-to-interference-plus-noise ratio (SINR) received by user $k\in \mathcal{K}$ from BS $n\in \mathcal{N}$ can be written as
\begin{equation}\label{eq1}
{{\text{SINR}}_{nk}}=\frac{{{p}_{n}}{{g}_{nk}}}{{{\sum }_{j\in \mathcal{N}\backslash \left\{ n \right\}}}{{p}_{j}}{{g}_{jk}}+{\sigma _{n}^{2}}},\ n\in \mathcal{N},
\end{equation}
where ${{p}_{n}}$ represents the transmit power of BS $n$; ${{g}_{nk}}$ denotes the channel gain between BS $n$ and user $k$; ${\sigma _{n}^{2}}$ is the noise power of BS $n$. Then, the achievable rate of user $k$ from BS $n$ can be represented as ${{r}_{nk}}={\log}_{2} \left( 1+{{{\text{SINR}}_{nk}}} \right)$.
\par
To proceed, we need to give the following definitions.
\par
\noindent
\emph{\textbf{Definition 1:}} The effective load of BS $n$ is represented as ${{y}_{n}}=\sum\nolimits_{k\in \mathcal{K}}{{{x}_{nk}}}$, where ${{x}_{nk}}$ is an association indicator, i.e., ${{x}_{nk}}=1$ when user $k$ is associated with BS $n$, 0 otherwise.
\par
\noindent
\emph{\textbf{Definition 2:}} If ${{y}_{n}}$ users are connected to BS $n$, the effective rate (long-term rate) of user $k$ associated with BS $n$ is given by ${{R}_{nk}}={{{r}_{nk}}}/{{{y}_{n}}}$.
\par
It is noteworthy that the definition 2 has three kinds of interpretations. In the first type, users associated with some BS are served by employing round-robin scheduling. In the second type, users associated with some BS perform equal bandwidth allocation. In the last type, ${{R}_{nk}}$ can be interpreted as load efficiency. Since the parameter ${{R}_{nk}}$ is closely related with achievable rate and load (number of associated users), it can be regarded as a key association factor for load balancing in HCNs. Unlike maximal (achievable) rate association, the association scheme based on the parameter ${{R}_{nk}}$ can reflect both the received signal strength and the load level.
\section{Association Schemes}
Next, we will give detailed descriptions for the two association schemes involved in this paper. Considering the characteristic of effective rate, we can take it as a key factor for the design of association algorithm. Specially, we design two association schemes that maximize the system throughput (sum of effective rates). In these two schemes, the one just performs user association, but the other jointly considers user association and power control. Note that the former refers to scheme UAMSER, and the latter refers to scheme JUAPCMSER.
\subsection{Scheme UAMSER}
To maximize the system throughput and balance the network loads among different BSs, we optimize such a problem that maximizes the sum of effective rates, and thus give the scheme UAMSER. Mathematically, it can be formulated as
\begin{equation}\label{eq2}
\begin{split}
\underset{\boldsymbol{x}}{\mathop{\max }}\,&\ \sum\limits_{n\in \mathcal{N}}{\sum\limits_{k\in \mathcal{K}}{{{x}_{nk}}{{R}_{nk}}}} \\
\text{s.t.}&\ \sum\limits_{n\in \mathcal{N}}{{{x}_{nk}}}=1,\forall k\in \mathcal{K}, \\
 &{{x}_{nk}}\in \left\{ 0,1 \right\},\forall n\in \mathcal{N},\forall k\in \mathcal{K}, \\
\end{split}
\end{equation}
where $\boldsymbol{x}=\left\{ {{x}_{nk}},\forall n\in \mathcal{N},\forall k\in \mathcal{K} \right\}$ and the first constraint shows that some user can be associated with only one BS.
\par
Seen from the association objective in the problem \eqref{eq2}, we can know that its every element is closely related to the load of some BS and the achievable rate of some user. In order to maximize the association objective, the users don't always select the BSs with high achievable rates. Although these BSs have high achievable rates, they may not provide sufficient resources for associated users when they are in an overloaded state. To achieve the high user experiences, some users will not select these overloaded BSs, but they are in favour of the nearest underloaded BSs. Evidently, the scheme that maximizes the sum of effective rates can relatively balance network loads among different BSs.
\par
Through a direct observation, we can easily find that the formulated problem is in a non-convex form and hard to tackle. To achieve its global optimal solution, we may need to search all possible combinations of user associations. However, it may be undesirable in the practical system, especially in the large-scale system. To effectively solve this problem, we try to design an algorithm by obeying the rules in the literatures \cite{23,24}.
\par
It is easy to find that the problem \eqref{eq2} can be converted into
\begin{equation}\label{eq3}
\begin{split}
  \underset{\boldsymbol{x},\boldsymbol{\omega  }}{\mathop{\max }}\,&\ \sum\limits_{n\in \mathcal{N}}{\sum\limits_{k\in \mathcal{K}}{{{x}_{nk}}{{\omega  }_{nk}}}} \\
 \text{s.t. }&\sum\limits_{n\in \mathcal{N}}{{{x}_{nk}}}=1,\forall k\in \mathcal{K}, \\
 &{{r}_{nk}}\ge {{\omega  }_{nk}}\sum\limits_{i\in \mathcal{K}}{{{x}_{ni}}},\forall n\in \mathcal{N},\forall k\in \mathcal{K}, \\
 &{{x}_{nk}}\in \left\{ 0,1 \right\},\forall n\in \mathcal{N},\forall k\in \mathcal{K}, \\
\end{split}
\end{equation}
\par
To meet the demand of algorithm design, i.e., avoid the case ``/0", we make some change for the second constraint in the problem \eqref{eq3}. Specially,
\begin{equation}\label{eq4}
\begin{split}
  \underset{\boldsymbol{x},\boldsymbol{\omega }}{\mathop{\max }}\,&\ \sum\limits_{n\in \mathcal{N}}{\sum\limits_{k\in \mathcal{K}}{{{x}_{nk}}{{\omega }_{nk}}}} \\
 \text{s.t. }&\sum\limits_{n\in \mathcal{N}}{{{x}_{nk}}}=1,\forall k\in \mathcal{K}, \\
 &{{r}_{nk}}\ge {{\omega }_{nk}}\left( 1+\sum\limits_{i\in \mathcal{K}}{{{x}_{ni}}} \right),\forall n\in \mathcal{N},\forall k\in \mathcal{K}, \\
 &{{x}_{nk}}\in \left\{ 0,1 \right\},\forall n\in \mathcal{N},\forall k\in \mathcal{K}, \\
\end{split}
\end{equation}
\par
Evidently, the second constraint in the problem \eqref{eq3} should be met when the one in the problem \eqref{eq4} is met, and the problem \eqref{eq4} achieves a lower bound of problem \eqref{eq3}.
\par
Similar to works \cite{23,24}, the problem \eqref{eq4} can be changed into a tractable form according to the following theorem.
\par
\noindent
\emph{\textbf{Theorem 1:}} If $\left( \boldsymbol{\bar{x}},\boldsymbol{\bar{\omega }} \right)$ is the solution of problem \eqref{eq4}, then there exist $\boldsymbol{\bar{\mu}}$, such that $\boldsymbol{\bar{x}}$ satisfies the Karush-Kuhn-Tucker (KKT) conditions \cite{25} of the following problem for $\boldsymbol{\mu}=\boldsymbol{\bar{\mu}}$ and $\boldsymbol{\omega}=\boldsymbol{\bar{\omega}}$.
\begin{equation}\label{eq5}
\begin{split}
  \underset{\boldsymbol{x}}{\mathop{\max }}\,&\ F\left( \boldsymbol{x} \right)=\sum\limits_{n\in \mathcal{N}}{\sum\limits_{k\in \mathcal{K}}{{{x}_{nk}}\left\{ {{\omega }_{nk}}-\sum\limits_{i\in \mathcal{K}}{{{\mu }_{ni}}{{\omega }_{ni}}} \right\}}} \\
 \text{s.t. }& \sum\limits_{n\in \mathcal{N}}{{{x}_{nk}}}=1,\forall k\in \mathcal{K}, \\
 &{{x}_{nk}}\in \left\{ 0,1 \right\},\forall n\in \mathcal{N},\forall k\in \mathcal{K}, \\
\end{split}
\end{equation}
\par
In addition, $\boldsymbol{\bar{x}}$ also satisfies the following system equations for $\boldsymbol{\mu}=\boldsymbol{\bar{\mu}}$ and $\boldsymbol{\omega}=\boldsymbol{\bar{\omega}}$.
\begin{equation}\label{eq6}
\left\{ \begin{split}
  {{\mu }_{nk}}=&\frac{{{x}_{nk}}}{1+\sum\limits_{i\in \mathcal{K}}{{{x}_{ni}}}},\forall n\in \mathcal{N},\forall k\in \mathcal{K},  \\
 {{\omega }_{nk}}=&\frac{{{r}_{nk}}}{1+\sum\nolimits_{i\in \mathcal{K}}{{{x}_{ni}}}},\forall n\in \mathcal{N},\forall k\in \mathcal{K},  \\
\end{split} \right.
\end{equation}
\par
On the contrary, if $\boldsymbol{\bar{x}}$ is the solution of problem \eqref{eq5} and satisfies the mentioned-above system equations for $\boldsymbol{\mu}=\boldsymbol{\bar{\mu}}$ and $\boldsymbol{\omega}=\boldsymbol{\bar{\omega}}$, $\left( \boldsymbol{\bar{x}},\boldsymbol{\bar{\omega }} \right)$ also satisfies the KKT conditions of problem \eqref{eq4}.
\par
\emph{Proof:}  Introducing Lagrangian multipliers $\boldsymbol{\mu }=\left\{ {{\mu }_{nk}},\forall n\in \mathcal{N},\forall k\in \mathcal{K} \right\}$ for the second constraint of problem \eqref{eq4}. Then, the Lagrangian function with respect to this constraint is
\begin{equation}\label{eq7}
\begin{split}
\mathcal{L}\left( \boldsymbol{x},\boldsymbol{\omega },\boldsymbol{\mu } \right)=&\sum\limits_{n\in \mathcal{N}}{\sum\limits_{k\in \mathcal{K}}{{{\mu }_{nk}}\left( {{r}_{nk}}-{{\omega }_{nk}}-{{\omega }_{nk}}\sum\limits_{i\in \mathcal{K}}{{{x}_{ni}}} \right)}}\\
&+\sum\limits_{n\in \mathcal{N}}{\sum\limits_{k\in \mathcal{K}}{{{x}_{nk}}{{\omega }_{nk}}}}.\\
\end{split}
\end{equation}
\par
Since $\left( \boldsymbol{\bar{x}},\boldsymbol{\bar{\omega }} \right)$ is the solution of problem \eqref{eq4}, there exist $\boldsymbol{\bar{\mu}}$ for any $n$ and $k$ such that partial KKT conditions of problem \eqref{eq4} are as follows
\begin{equation}\label{eq8}
\frac{\partial \mathcal{L}}{\partial {{\omega }_{nk}}}={{\bar{x}}_{nk}}-{{\bar{\mu }}_{nk}}\left( 1+\sum\limits_{i\in \mathcal{K}}{{{{\bar{x}}}_{ni}}} \right)=0,
\end{equation}
\begin{equation}\label{eq9}
{\bar{\mu }_{nk}}\frac{\partial \mathcal{L}}{\partial {{\mu }_{nk}}}={{\bar{\mu }}_{nk}}\left( {{r}_{nk}}-{\bar{\omega }_{nk}}-{\bar{\omega }_{nk}}\sum\limits_{i\in \mathcal{K}}{{{x}_{ni}}} \right)=0.
\end{equation}
\par
Considering that $1+\sum\nolimits_{i\in \mathcal{K}}{{{{\bar{x}}}_{ni}}}>0$ for any $n$, we can achieve the following results according to the equations \eqref{eq8} and \eqref{eq9}.
\begin{equation}\label{eq10}
{{\bar{\mu }}_{nk}}=\frac{{{{\bar{x}}}_{nk}}}{1+\sum\limits_{i\in \mathcal{K}}{{{{\bar{x}}}_{ni}}}},\forall n\in \mathcal{N},\forall k\in \mathcal{K},
\end{equation}
\begin{equation}\label{eq11}
{{\bar{\omega }}_{nk}}=\frac{{{r}_{nk}}}{1+\sum\nolimits_{i\in \mathcal{K}}{{{{\bar{x}}}_{ni}}}},\forall n\in \mathcal{N},\forall k\in \mathcal{K}.
\end{equation}
It is easy to find that system  equations \eqref{eq8} and \eqref{eq9} are the KKT conditions of the following problem for $\boldsymbol{\mu}=\boldsymbol{\bar{\mu}}$ and $\boldsymbol{\omega}=\boldsymbol{\bar{\omega}}$.
\begin{equation}\label{eq12}
\begin{split}
  \underset{\boldsymbol{x}}{\mathop{\max }}\,&\ \sum\limits_{n\in \mathcal{N}}{\sum\limits_{k\in \mathcal{K}}{\left\{ {{x}_{nk}}{{\omega }_{nk}}+{{\mu }_{nk}}\left( {{r}_{nk}}-{{\omega }_{nk}}-{{\omega }_{nk}}\sum\limits_{i\in \mathcal{K}}{{{x}_{ni}}} \right) \right\}}}\\
 \text{s.t. }& \sum\limits_{n\in \mathcal{N}}{{{x}_{nk}}}=1,\forall k\in \mathcal{K}, \\
 &{{x}_{nk}}\in \left\{ 0,1 \right\},\forall n\in \mathcal{N},\forall k\in \mathcal{K}, \\
\end{split}
\end{equation}
\par
Evidently, the problem \eqref{eq12} can be simplified into the problem \eqref{eq5}. Thus, the first conclusion of Theorem 1 holds. Similarly, the contrary conclusion can be easily proved.
\par
Theorem 1 shows that the solution of problem \eqref{eq4} can be found among the solutions of problem \eqref{eq5}, which satisfies system equation \eqref{eq6}. In addition, such a solution should be global solution of problem \eqref{eq4} if it is unique \cite{23,24}.
\par
Through direct observation from the problem \eqref{eq5}, it is easy to know that its solution can be found according to the following rule.
\begin{equation}\label{eq13}
{{n}^{*}}=\arg \underset{n\in \mathcal{N}}{\mathop{\max }}\,\left\{ {{\omega }_{nk}}-\sum\limits_{i\in \mathcal{K}}{{{\mu }_{ni}}{{\omega }_{ni}}} \right\},\forall k\in \mathcal{K},
\end{equation}
The rule \eqref{eq13} shows that any user $k$ selects some BS $n^*$ to maximize the obtained utility ${{\omega }_{{n^{*}}k}}-\sum\nolimits_{i\in \mathcal{K}}{{{\mu }_{{n^{*}}i}}{{\omega }_{{n^{*}}i}}}$. In other words, any user $k$ selects some BS $n^*$ when the utility ${{\omega }_{{n^{*}}k}}-\sum\nolimits_{i\in \mathcal{K}}{{{\mu }_{{n^{*}}i}}{{\omega }_{{n^{*}}i}}}$ is the maximum among all possible associations.
\par
Based on aforementioned analyses, a one-layer iterative algorithm to solve the problem \eqref{eq4} can be easily given. The detailed descriptions can be found in Algorithm 1, where $\boldsymbol{\mu}$ and $\boldsymbol{\omega}$ are updated via Newton-like method; $\boldsymbol{x}$ is decided according to the rule \eqref{eq13}; moreover, we give the definitions of some functions for any $n$ and $k$ as follows:
\begin{equation}\label{eq14}
{{\phi }_{nk}}\left( {{\mu }_{nk}} \right)={{\mu }_{nk}}\left( 1+\sum\limits_{i\in \mathcal{K}}{{{x}_{ni}}} \right)-{{x}_{nk}},
\end{equation}
\begin{equation}\label{eq15}
{{\varphi }_{nk}}\left( {{\omega }_{nk}} \right)={{\omega }_{nk}}\left( 1+\sum\limits_{i\in \mathcal{K}}{{{x}_{ni}}} \right)-{{r}_{nk}},
\end{equation}
\begin{equation}\label{eq16}
{{\chi }_{n}}=\frac{1}{1+\sum\nolimits_{i\in \mathcal{K}}{{{{{x}}}_{ni}}}}.
\end{equation}

\begin{table}[]
\centering
\begin{tabular}{ll}
\toprule[1pt]
\textbf{Algorithm 1: UAMSER} \\ \midrule[0.5pt]
1: \textbf{Initialization:} Set $\xi =0.5$, $\varepsilon ={{10}^{-3}}$, ${{t}_{1}}=1$, and take arbitrarily ${{\boldsymbol{x}}^{t}}$ \\
\ \ \ that satisfies the constraints of problem \eqref{eq5}. Let\\
\ \ \ \ \ \ \ \ \ \ \ \ $\mu _{nk}^{{t}_{1}}={x_{nk}^{{t}_{1}}}/{\left( 1+\sum\limits_{i\in \mathcal{K}}{x_{ni}^{{t}_{1}}} \right)}\;,\forall n\in \mathcal{N},\forall k\in \mathcal{K},$ \\
\ \ \ \ \ \ \ \ \ \ \ \ $\omega _{nk}^{{t}_{1}}={{{r}_{nk}}}/{\left( 1+\sum\limits_{i\in \mathcal{K}}{x_{ni}^{{t}_{1}}} \right)}\;,\forall n\in \mathcal{N},\forall k\in \mathcal{K},$ \\
2: \textbf{Repeat (Main Loop)}\\
3: \ \ \ \ Any user selects some BS according to the rule \eqref{eq14}.\\
4: \ \ \ \ If the following conditions are satisifed, then stop the algorithm. \\
\ \ \ \ \ \ \ Otherwise, go to step 5.\\
\ \ \ \ \ \ \ \ \ \ \ \ $\mu _{nk}^{{t}_{1}}\left( 1+\sum\limits_{i\in \mathcal{K}}{x_{ni}^{{t}_{1}}} \right)-x_{nk}^{{t}_{1}}=0,\forall n\in \mathcal{N},\forall k\in \mathcal{K},$\\
\ \ \ \ \ \ \ \ \ \ \ \ $\omega _{nk}^{{t}_{1}}\left( 1+\sum\limits_{i\in \mathcal{K}}{x_{ni}^{{t}_{1}}} \right)-{{r}_{nk}}=0,\forall n\in \mathcal{N},\forall k\in \mathcal{K},$\\
5: \ \ \ \ Find the smallest $m$ among $m\in \left\{ 0,1,2,\cdots  \right\}$ satisfying\\
\ \ \ \ \ \ \ \ \ \ \ \ $\sum\limits_{n\in \mathcal{N}}{\sum\limits_{k\in \mathcal{K}}{{{\left| {{\phi }_{nk}}\left( \mu _{nk}^{{t}_{1}}-{{\xi }^{m}}{{\chi }_{n}}{{\phi }_{nk}}\left( \mu _{nk}^{{t}_{1}} \right) \right) \right|}^{2}}}}$ \\
\ \ \ \ \ \ \ \ \ \ \ \ \ \ \ +$\sum\limits_{n\in \mathcal{N}}{\sum\limits_{k\in \mathcal{K}}{{{\left| {{\varphi }_{nk}}\left( \omega _{nk}^{{t}_{1}}-{{\xi }^{m}}{{\chi }_{n}}{{\varphi }_{nk}}\left( \omega _{nk}^{{t}_{1}} \right) \right) \right|}^{2}}}}$ \\
\ \ \ \ \ \ \ \ \ \ \ \ $\le \left( 1-\varepsilon {{\xi }^{m}} \right)\sum\limits_{n\in \mathcal{N}}{\sum\limits_{k\in \mathcal{K}}{\left\{ {{\left| {{\phi }_{nk}}\left( \mu _{nk}^{{t}_{1}} \right) \right|}^{2}}+{{\left| {{\varphi }_{nk}}\left( \omega _{nk}^{{t}_{1}} \right) \right|}^{2}} \right\}}}$ \\
6: \ \ \ \ Update $\boldsymbol{\mu }$ and $\boldsymbol{\omega }$ according Newton-like method:\\
\ \ \ \ \ \ \ \ \ \ \ \ $\mu _{nk}^{{{t}_{1}}+1}=\mu _{nk}^{{t}_{1}}-{{\xi }^{m}}{{\chi }_{n}}{{\phi }_{nk}}\left( \mu _{nk}^{{t}_{1}} \right),\forall n\in \mathcal{N},\forall k\in \mathcal{K},$ \\
\ \ \ \ \ \ \ \ \ \ \ \ $\omega _{nk}^{{{t}_{1}}+1}=\omega _{nk}^{{t}_{1}}-{{\xi }^{m}}{{\chi }_{n}}{{\varphi }_{nk}}\left( \omega _{nk}^{{t}_{1}} \right),\forall n\in \mathcal{N},\forall k\in \mathcal{K},$ \\
7: \ \ \ \ Normalize $\mu _{nk}^{{{t}_{1}}+1}$, i.e.,  $\mu _{nk}^{{{t}_{1}}+1}={\mu _{nk}^{{{t}_{1}}+1}}/{\sum\nolimits_{n\in \mathcal{N}}{\sum\nolimits_{k\in \mathcal{K}}{\mu _{nk}^{{{t}_{1}}+1}}}}$.\\
8: \ \ \ \ ${{t}_{1}}={{t}_{1}}+1$.\\
9: \textbf{Until} $F\left( \boldsymbol{x}\right)$ converges or ${{t}_{1}}={{T}_{1}}$.  \\ \bottomrule[0.5pt]
\end{tabular}
\label{alg1}
\end{table}
\par
In Algorithm 1, ${{t}_{1}}$ is the iteration index, ${{T}_{1}}$ represents the maximal number of iterations and the step 7 normalizes multipliers to ensure that the Lagrangian function \eqref{eq7} is bounded.
\par
Next, we will investigate the convergence of Algorithm 1.
\par
\noindent
\emph{\textbf{Theorem 2:}} The Algorithm 1 is guaranteed to converge.
\par
\emph{Proof}: In the steps 5 and 6 of Algorithm 1, $\boldsymbol{\mu}$ and $\boldsymbol{\omega}$ are updated using a Newton-like method, which has a linear convergence rate. When ${{\xi }^{m}}=1$, the update of $\boldsymbol{\mu}$ and $\boldsymbol{\omega}$ reduces to the Newton method who has a quadratic convergence rate. Specially, the convergence of Algorithm 1 can be proven by employing a similar method used in work \cite{23}.
\subsection{Scheme JUAPCMSER}
To maximize the system throughput and balance the network loads among different BSs, we also optimize such a problem that maximizes the sum of effective rates. Moreover, in order to mitigate the interference and reduce the energy consumption, we further consider power control in the user association and thus give the scheme JUAPCMSER. Mathematically, it can be formulated as
\begin{equation}\label{eq17}
\begin{split}
   \underset{\boldsymbol{p},\boldsymbol{x}}{\mathop{\max }}\,\ &F\left( \boldsymbol{p},\boldsymbol{x} \right)=\sum\limits_{n\in \mathcal{N}}{\sum\limits_{k\in \mathcal{K}}{{{x}_{nk}}{{R}_{nk}}\left( \boldsymbol{p},\boldsymbol{x} \right)}} \\
  \text{s.t. }&\sum\limits_{n\in \mathcal{N}}{{{x}_{nk}}}=1,\forall k\in \mathcal{K}, \\
 &0\le {{p}_{n}}\le p_{n}^{\text{max} },\forall n\in \mathcal{N}, \\
 & {{x}_{nk}}\in \left\{ 0,1 \right\},\forall n\in \mathcal{N},\forall k\in \mathcal{K}, \\
\end{split}
\end{equation}
where $\boldsymbol{p}=\left\{ {{p}_{n}}, \forall n\in \mathcal{N} \right\}$; ${p_{n}^{\text{max} }}$ is the maximal (allowed) transmit power of BS $n$.
\par
Since the objective function $F\left( \boldsymbol{p},\boldsymbol{x} \right)$ of problem \eqref{eq17} is non-convex with respect to transmit power $\boldsymbol{p}$ and is also tightly coupled with the integer association index $\boldsymbol{x}$, this problem is a mixed-integer and nonlinear programming. Compared with the problem \eqref{eq2}, the problem \eqref{eq17} has a more higher computation complexity. To achieve its global optimal solution, we need to fully search the feasible power space with a small granularity along with all possible combinations of user associations. Thus, even for a centralized system, it may be infeasible to solve the problem \eqref{eq17} at each association slot.
\par
Now, we present a two-layer iterative algorithm to achieve a sub-optimal solution of problem \eqref{eq17}, which alternatively optimizes transmit powers of BSs and association indices of users. Specially, the outer layer performs user association under fixed transmit powers, and the inner layer updates the transmit powers under fixed association indices.
\par
For any given feasible power $\boldsymbol{p}$, we can easily obtain the simplified form of problem \eqref{eq17}, i.e., problem \eqref{eq2}. Thus, we can solve the problem \eqref{eq17} with fixed transmit power using Algorithm 1. In addition, for any given user association index $\boldsymbol{x}$, the problem \eqref{eq17} can be simplified into the following power control problem:
\begin{equation}\label{eq18}
\begin{split}
  \underset{\boldsymbol{p}}{\mathop{\max }}\,&\ F\left( \boldsymbol{p} \right)=\sum\limits_{n\in \mathcal{N}}{\sum\limits_{k\in \mathcal{K}}{{{x}_{nk}}{{R}_{nk}}\left( \boldsymbol{p} \right)}} \\
 \text{s.t. }&0\le {{p}_{n}}\le p_{n}^{\text{max} },\forall n\in \mathcal{N}, \\
\end{split}
\end{equation}
\par
According to the relation between effective and achievable rates, the problem \eqref{eq18} can be converted into
\begin{equation}\label{eq19}
\begin{split}
  \underset{\boldsymbol{p}}{\mathop{\max }}\,&\ F\left( \boldsymbol{p} \right)=\sum\limits_{n\in \mathcal{N}}{\sum\limits_{k\in \mathcal{K}}{ {{\eta}_{nk}}\log \left( 1+{{\text{SINR}}_{nk}}\left( \boldsymbol{p} \right) \right)}} \\
 \text{s.t. }&0\le {{p}_{n}}\le p_{n}^{\text{max} },\forall n\in \mathcal{N}, \\
\end{split}
\end{equation}
where ${{\eta}_{nk}}={{{x}_{nk}}}/{\sum\nolimits_{i\in \mathcal{K}}{{{x}_{ni}}}}$. To meet the demand of algorithm design, we let ${{\eta}_{nk}}=0$ when ${\sum\nolimits_{i\in \mathcal{K}}{{{x}_{ni}}}}=0$. In other words, the throughput of some BS should be 0 when no user is served by it.
\par
Considering that the objective function of problem \eqref{eq19} is in a non-convex form, we need to make some changes to achieve its convex form. Similar to works \cite{26,27,28}, we let $\log \left( 1+{{\text{SINR}}_{nk}}\left( \boldsymbol{p} \right) \right)\approx \log {{\text{SINR}}_{nk}}\left( \boldsymbol{p} \right)$ and ${{\bar{p}}_{n}}=\log {{p}_{n}}$. It can be easily found that the first operation lets us have a lower bound of original problem. According to these changes, we have the following convex optimization problem:
\begin{equation}\label{eq20}
\begin{split}
  \underset{{\boldsymbol{\bar{p}}}}{\mathop{\max }}\,&\ F\left( {\boldsymbol{\bar{p}}} \right)=\sum\limits_{n\in \mathcal{N}}{\sum\limits_{k\in \mathcal{K}}{{{\eta}_{nk}}\log\,{{\overline{\text{SINR}}}_{nk}}\left( {\boldsymbol{\bar{p}}} \right)}} \\
  \text{s.t. }&{{{\bar{p}}}_{n}}\le \log p_{n}^{\text{max} },\forall n\in \mathcal{N}, \\
\end{split}
\end{equation}
where $\boldsymbol{\bar{p}}=\left\{ {{{\bar{p}}}_{n}},\forall n\in \mathcal{N} \right\}$ and
\begin{equation}\label{eq21}
{{\overline{\text{SINR}}}_{nk}}={{\overline{\text{SINR}}}_{nk}}\left( {{\boldsymbol{\bar{p}}}} \right)=\frac{{{e}^{{{{\bar{p}}}_{n}}}}{{g}_{nk}}}{{{\sum }_{j\in \mathcal{N}\backslash \left\{ n \right\}}}{{e}^{{{{\bar{p}}}_{j}}}}{{g}_{jk}}+\sigma _{n}^{2}}.
\end{equation}
\par
According to the extreme value principle, we can easily know that the optimal transmit power should meet the condition: ${\partial F}/{\partial {{{\bar{p}}}_{m}}}=0$ for all $m$. Thus, we have
\begin{equation}\label{eq22}
{{e}^{{{{\bar{p}}}_{m}}}}=\frac{\sum\limits_{k\in \mathcal{K}}{{{\eta}_{mk}}}}{{{\ell }_{m}}\left( {\boldsymbol{\bar{p}}} \right)},\forall m\in \mathcal{N},
\end{equation}
where
\begin{equation}\label{eq23}
{{\ell }_{m}}\left( {\boldsymbol{\bar{p}}} \right)=\sum\limits_{n\in \mathcal{N}\backslash \left\{ m \right\}}{\sum\limits_{k\in \mathcal{K}}{\frac{{{\eta}_{nk}}{{g}_{mk}}}{{{\sum }_{j\in \mathcal{N}\backslash \left\{ n \right\}}}{{e}^{{{{\bar{p}}}_{j}}}}{{g}_{jk}}+{\sigma _{n}^{2}}}}}.\\
\end{equation}
\par
Since $\boldsymbol{a}\preccurlyeq \boldsymbol{\bar{p}}\preccurlyeq \boldsymbol{b}$, we state the following lemma to show that the KKT conditions of \eqref{eq19} are equivalent to projecting \eqref{eq21} to $\left[ \log {{a}_{m}},\log {{b}_{m}} \right]$ for all $m$.
\par
\noindent
\emph{\textbf{Lemma 1 \cite{28,29}:}} There is an optimization problem ${{\min }_{\boldsymbol{a}\preccurlyeq \boldsymbol{z}\preccurlyeq \boldsymbol{b}}}\ f\left( \boldsymbol{z} \right)$. Then, its KKT conditions are equivalent to the condition ${{P}_{[ \boldsymbol{a},\boldsymbol{b} ]}}{\partial f}/{\partial \boldsymbol{z}}\;=0$, where ${{P}_{[ \boldsymbol{a},\boldsymbol{b} ]}}\boldsymbol{\chi }$ is the projection of vector $\boldsymbol{\chi }$ onto the box $[ \boldsymbol{a},\boldsymbol{b} ]$ defined by ${{( {{P}_{[ \boldsymbol{a},\boldsymbol{b} ]}}\boldsymbol{\chi } )}_{k}}=\min \left\{ 0,{{\chi }_{k}} \right\}$ if ${{z}_{k}}={{a}_{k}}$, ${{( {{P}_{[ \boldsymbol{a},\boldsymbol{b} ]}}\boldsymbol{\chi } )}_{k}}={{\chi }_{k}}$ if ${{z}_{k}}\in [ {{{a}}_{k}},{{{b}}_{k}} ]$ and ${{( {{P}_{[ \boldsymbol{a},\boldsymbol{b} ]}}\boldsymbol{\chi } )}_{k}}=\max \left\{ 0,{{\chi }_{k}} \right\}$ if ${{z}_{k}}={{b}_{k}}$.
\par
 Applying Lemma 1 to the problem \eqref{eq5} in the $\boldsymbol{\bar{p}}$ domain $[ \log \boldsymbol{a},\log \boldsymbol{b} ]$ and then converting it back to the $\boldsymbol{p}$  domain $[ \boldsymbol{a},\boldsymbol{b} ]$, we can deduce
\begin{equation}\label{eq24}
p_{m}^{t+1}={{I}_{m}}\left( {{\boldsymbol{p}}^{t}} \right)={{\left[ \frac{\sum\limits_{k\in \mathcal{K}}{{{\eta}_{mk}}}}{{{\hbar }_{m}}\left( {{\boldsymbol{p}}^{t}} \right)} \right]}^{\log p_{n}^{\text{max} }}},\forall m\in \mathcal{N},
\end{equation}
where $t$ is iteration index and
\begin{equation}\label{eq25}
\begin{split}
  {{\hbar }_{m}}\left( {{\boldsymbol{p}}^{t}} \right) =&\sum\limits_{n\in \mathcal{N}\backslash \left\{ m \right\}}{\sum\limits_{k\in \mathcal{K}}{\frac{{{\eta}_{nk}}{{g}_{mk}}}{{{\sum }_{j\in \mathcal{N}\backslash \left\{ n \right\}}}p_{j}^{t}{{g}_{jk}}+{\sigma _{n}^{2}}}}}\\
 =&\sum\limits_{n\in \mathcal{N}\backslash \left\{ m \right\}}{\sum\limits_{k\in \mathcal{K}}{{{g}_{mk}}{{\Gamma }_{nk}}\left( {{\boldsymbol{p}}^{t}} \right)}} .\\
 \end{split}
\end{equation}
\par
Now, we give a detailed procedure for joint user association and power control, which is listed in Algorithm 2. In this algorithm, ${{T}_{2}}$ and ${{T}_{3}}$ are the maximal number of iterations of inner loop and the one of outer loop respectively, and ${{t}_{2}}$ and ${{t}_{3}}$ are the iteration indices of inner and outer loops respectively.
\par
\begin{table}[h]
\centering
\begin{tabular}{ll}
\toprule[1pt]
\textbf{Algorithm 2} JUAPCMSER\\ \midrule[0.5pt]
1: \textbf{Initialization:} ${{t}_{3}}=1$. \\
2: \textbf{Repeat (Outer Loop)}\\
3:\ \ \ \ Run the Algorithm 1 to obtain association index $\boldsymbol{x}$.\\
4:\ \ \ \ \textbf{Initialization:} ${{t}_{2}}=1$ and ${\boldsymbol{p}}^{{t}_{2}}=\boldsymbol{p}^{\text{max} }$. \\
5:\ \ \ \  \textbf{Repeat (Inner Loop)}\\
6:\ \ \ \ \ \ \ \ Update the transmit power ${{\boldsymbol{p}}^{{t}_{2}+1}}$ using \eqref{eq24}.\\
7:\ \ \ \ \ \ \ \ ${{t}_{2}}={{t}_{2}}+1$.\\
8:\ \ \ \ \textbf{Until} ${\boldsymbol{p}}$ converges or ${{t}_{2}}={{T}_{2}}$.\\
9:\ \ \ \ ${{t}_{3}}={{t}_{3}}+1$.\\
10:\textbf{Until} $F\left( \boldsymbol{p},\boldsymbol{x} \right)$ converge or ${{t}_{3}}={{T}_{3}}$\\ \bottomrule[0.5pt]
\end{tabular}
\label{tab1}
\end{table}
\par
Next, we give the proof of convergence of Algorithm 2. Considering the convergence of association procedure of Algorithm 2 has been proven in Theorem 2, we just need to prove the convergence of power control algorithm (inner loop) of Algorithm 2. After that, we can easily know that the Algorithm 2 should converge. To prove the convergence of inner loop of Algorithm 2, we just need to show that we can find a stationary point ${{\boldsymbol{p}}^{*}}$ of the power update procedure for any given association index $\boldsymbol{x}$. To this end, we need to introduce the definition of a two-sided scalable (2.s.s.) function as follows.
\par
\noindent
\emph{\textbf{Definition 3:}} A power update function (puf) $\boldsymbol{f}\left( \boldsymbol{p} \right)={{\left[ {{\operatorname{f}}_{1}}\left( \boldsymbol{p} \right),\cdots ,{{\operatorname{f}}_{N}}\left( \boldsymbol{p} \right) \right]}^{T}}$ is two-sided scalable (2.s.s.) with respect to $\boldsymbol{p}=\left\{ {{p}_{n}},\forall n\in \mathcal{N} \right\}$ if for all $a>1$ and any power vector $\boldsymbol{\bar{p}}=\left\{ {{{\bar{p}}}_{n}},\forall n\in \mathcal{N} \right\}$ satisfying $\left( 1/a \right)\boldsymbol{p}\preccurlyeq \boldsymbol{\bar{p}}\preccurlyeq a\boldsymbol{p}$, we have
\begin{equation}\label{eq26}
\left( 1/a \right){{\operatorname{f}}_{n}}\left( \boldsymbol{p} \right)\le {{\operatorname{f}}_{n}}\left( {\boldsymbol{\bar{p}}} \right)\le a{{\operatorname{f}}_{n}}\left( \boldsymbol{p} \right),\ \forall n\in \mathcal{N}.
\end{equation}
\par
To prove the convergence of inner iterative loop (power control algorithm) using a 2.s.s. function approach, we recall the convergence results for any power control algorithm that employs a bounded 2.s.s. puf in the following Lemma \cite{30}.
\par
\noindent
\emph{\textbf{Lemma 2:}} Assume that $\boldsymbol{f}\left( \boldsymbol{p} \right)$ is a 2.s.s. function, whose element ${{f}_{n}}\left( \boldsymbol{p} \right),\forall n\in \mathcal{N}$ is bounded by zero and ${ f_{n}^{\max }}$, i.e., $0\le {{f}_{n}}\left( \boldsymbol{p} \right)\le { f_{n}^{\max }}$. Consider the corresponding power update $p_{n}^{t+1}={{f}_{n}}\left( {{\boldsymbol{p}}^{t}} \right)$. Then, we have the following results:
\begin{enumerate}[1)]
\item  Given a transmit power ${{\boldsymbol{p}}^{*}}$, the puf $\boldsymbol{f}\left( \boldsymbol{p} \right)$ has a unique fixed point that satisfies ${{\boldsymbol{p}}^{*}}=\boldsymbol{f}\left( {{\boldsymbol{p}}^{*}} \right)$;

\item  Given an arbitray initial power vector ${{\boldsymbol{p}}^{0}}$, the power control algorithm based on puf $\boldsymbol{f}\left( \boldsymbol{p} \right)$ converges to the unique and fixed point ${{\boldsymbol{p}}^{*}}$.

\end{enumerate}
\emph{Proof:} The results of this lemma have been established for the 2.s.s. puf in \cite{30}.
\par
\noindent
\emph{\textbf{Theorem 3:}} The inner loop (power control algorithm) of Algorithm 2 converges to the unique and fixed point.
\par
\emph{Proof}: To prove the convergence of inner loop of Algorithm 2, we first show that the puf $I\left( \boldsymbol{p} \right)={{\left[ {{I}_{1}}\left( \boldsymbol{p} \right),\ldots ,{{I}_{N}}\left( \boldsymbol{p} \right) \right]}^{T}}$ is a 2 s.s. function with respect to $\boldsymbol{p}$. Then, the convergence of inner loop of Algorithm 2 can be proved by employing the results of Lemma 2.
\par
We assume that $\left( 1/a \right)\boldsymbol{p}\preccurlyeq \boldsymbol{\bar{p}}\preccurlyeq a\boldsymbol{p}$, where $a>1$. Since ${{\eta}_{nk}}\ge 0$ for any $n$ and $k$, we can deduce
\begin{equation}\label{eq27}
\begin{split}
  {{\Gamma }_{nk}}\left( {\mathbf{\bar{p}}} \right)& =\frac{{{\eta }_{nk}}{{g}_{mk}}}{\sum\nolimits_{j\in \mathcal{N}\backslash \left\{ n \right\}}{{{{\bar{p}}}_{j}}{{g}_{jk}}}+\sigma _{n}^{2}} \\
 &\le \frac{{{\eta }_{nk}}{{g}_{mk}}}{\sum\nolimits_{j\in \mathcal{N}\backslash \left\{ n \right\}}{{{{p}_{j}}{{g}_{jk}}}/{a}\;}+\sigma _{n}^{2}} \\
 &\le \frac{{{\eta }_{nk}}{{g}_{mk}}}{\sum\nolimits_{j\in \mathcal{N}\backslash \left\{ n \right\}}{{{{p}_{j}}{{g}_{jk}}}/{a}\;}+{\sigma _{n}^{2}}/{a}\;} \\
 &=a{{\Gamma }_{nk}}\left( \mathbf{p} \right), \\
\end{split}
\end{equation}
\par
\noindent
\begin{equation}\label{eq28}
\begin{split}
  {{\Gamma }_{nk}}\left( {\mathbf{\bar{p}}} \right)& =\frac{{{\eta }_{nk}}{{g}_{mk}}}{\sum\nolimits_{j\in \mathcal{N}\backslash \left\{ n \right\}}{{{{\bar{p}}}_{j}}{{g}_{jk}}}+\sigma _{n}^{2}} \\
 &\ge \frac{{{\eta }_{nk}}{{g}_{mk}}}{\sum\nolimits_{j\in \mathcal{N}\backslash \left\{ n \right\}}{a{{p}_{j}}{{g}_{jk}}}+\sigma _{n}^{2}} \\
 &\ge \frac{{{\eta }_{nk}}{{g}_{mk}}}{\sum\nolimits_{j\in \mathcal{N}\backslash \left\{ n \right\}}{a{{p}_{j}}{{g}_{jk}}}+a\sigma _{n}^{2}} \\
 &=\left( {1}/{a}\; \right){{\Gamma }_{nk}}\left( \mathbf{p} \right). \\
\end{split}
\end{equation}
\par
\noindent
Similarly, we have
\begin{equation}\label{eq29}
\left( {1}/{a}\; \right){{\hbar }_{m}}\left( \boldsymbol{p} \right)\le {{\hbar }_{m}}\left( {\boldsymbol{\bar{p}}} \right)\le a{{\hbar }_{m}}\left( \boldsymbol{p} \right).
\end{equation}
\par
\noindent
Furthermore, we can easily obtain the following result.
\begin{equation}\label{eq30}
\left( 1/a \right){{I}_{m}}\left( \boldsymbol{p} \right)\le {{I}_{m}}\left( {\boldsymbol{\bar{p}}} \right)\le a{{I}_{m}}\left( \boldsymbol{p} \right).
\end{equation}
Evidently, the puf $I\left( \boldsymbol{p} \right)$ is a 2.s.s. function. According to Lemma 2, we know that the inner loop of Algorithm 2 converges to the unique and fixed point.
\par

\subsection{Complexity Analysis}
Since the term $\sum\nolimits_{i\in \mathcal{K}}{x_{ni}}$ can be calculated before performing the steps 1 and 4 of Algorithm 1, these steps have a complexity of $\mathcal{O}\left( NK \right)$. Considering that the step 5 of Algorithm 1 needs to find the smallest $m$, we can deduce that this step has a complexity of $\mathcal{O}\left( \left( m+1\right)NK \right)$. As for other steps of Algorithm 1, we can easily find that they have a complexity of $\mathcal{O}\left( NK \right)$. Thus, the computation complexity of Algorithm 1 is $\mathcal{O}\left( \left( m+1 \right){{T}_{1}}NK \right)$, where $m$ often takes a very small integer number. Unlike Algorithm 1, the Algorithm 2 alternatively optimizes the transmit powers of BSs and the association indices of users, and thus occupies a higher computation complexity than the former. In the inner loop (power control algorithm) of Algorithm 2, each BS updates its transmit power using equation \eqref{eq24} that results in a complexity of $\mathcal{O}\left( NK \right)$, and thus the total computation complexity is $\mathcal{O}\left( {{N}^{2}}K \right)$ for all BSs. Similar to Algorithm 1, the user association procedure of Algorithm 2 has a complexity of $\mathcal{O}\left( \left( m+1 \right){{T}_{1}}NK \right)$. In general, the complexity of Algorithm 2 is the maximum between $\mathcal{O}\left( \left( m+1 \right){{T}_{1}}{{T}_{3}}NK\right)$ and $\mathcal{O}\left( {{T}_{2}}{{T}_{3}}{{N}^{2}}K \right)$. According to the simulation results, we know that the inner and outer loops can converge in very small numbers of iterations, and thus ${{T}_{2}}$ and ${{T}_{3}}$ are very small. Evidently, these proposed algorithms can be well implemented in the practical system.

\section{Numerical results}
In HCNs, the location of MBSs is fixed to form a conventional cellular framework, while the PBSs and users are scattered into each macrocell in a relatively random manner. We assume that the inter-site distance between any two MBSs is 1000 m, the maximal transmit powers of MBS and PBS are 46 dBm and 30 dBm respectively, the circuit powers of MBS and PBS are 10 W and 0.1 W respectively, the coefficients of power amplifier of MBS and PBS are 4 and 2 respectively, the noise power spectral density is -174 dBm/Hz and the system bandwidth is 10 MHz. In HCNs, we adopt the pathloss models $l_{nk}=128.1+37.6\log 10\left( d_{nk} \right)$ and $l_{nk}=140.7+36.7\log 10\left( d_{nk} \right)$ for MBS and PBS respectively \cite{15}, where $d_{nk}$ is the distance between user $k$ and BS $n$. Meanwhile, we consider the log-normal shadowing with a standard deviation 8 dB in the propagation environment.
\par
In the simulation, we will compare the two proposed association schemes with others. The former includes association schemes UAMSER and JUAPCMSER, and the latter certainly includes maximal achievable rate association (MARA) and association with user fairness (AUF) \cite{12}. In the compared schemes, the scheme AUF is an offloading one, but it is not the case for scheme MARA. Considering that the schemes UAMSER and JUAPCMSER are closely related to the loads of BSs and the achievable rates of users, and thus can relatively balance the network loads among different BSs, we can regard them as offloading schemes. In order to show the load balancing gain and power control gain, we will investigate different association performances such as load balancing level, energy efficiency and cumulative distribution function (CDF) of effective rates. In addition, we also show the convergence of Algorithms.
\begin{figure}[!t]
\centering
\centerline{\includegraphics[width=3.8in]{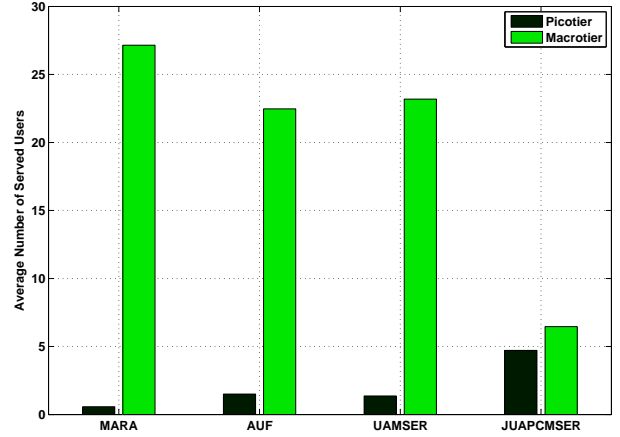}}
\caption{Load distributions among different network tiers for different association schemes.}
\label{fig2}
\end{figure}
\par
Fig. \ref{fig2} investigates the average numbers of users served by per tier for different association schemes, where the picotier consists of all PBSs and the macrotier is composed of all MBSs. In the scheme MARA, most users are attracted by MBSs according to the signal strength. Thus, the scheme MARA has more users associated with mcrotier than offloading schemes. Unlike scheme UAMSER, the scheme AUF enhances the user fairness while balancing network loads. Evidently, the enhancement of user fairness is beneficial to improving the load balancing level. Consequently, the scheme AUF has more users associated with picotier and fewer users associated with macrotier than scheme UAMSER. Since the power control of scheme JUAPCMSER narrows the gap between the transmit powers of MBS and PBS, it ensures that macrotier and picotier just have slightly different numbers of served users.
\par
To accurately measure the load balancing level of the network, we introduce a Jain's fairness index \cite{31} as a load balancing index, and give it by
\begin{equation}\label{eq31}
\gamma =\frac{{{\left( \sum\nolimits_{n\in \mathcal{N}}{{{y}_{n}}} \right)}^{2}}}{N\sum\nolimits_{n\in \mathcal{N}}{y_{n}^{2}}},
\end{equation}
where $\sum\nolimits_{k\in \mathcal{K}}{{{x}_{nk}}}={{y}_{n}}$ represents the load of BS ${n}$. A larger $\gamma$, taking value from the interval $\left[ \frac{1}{N},1 \right]$,  means a more balanced load distribution.
\begin{figure}[!t]
\centering
\centerline{\includegraphics[width=3.8in]{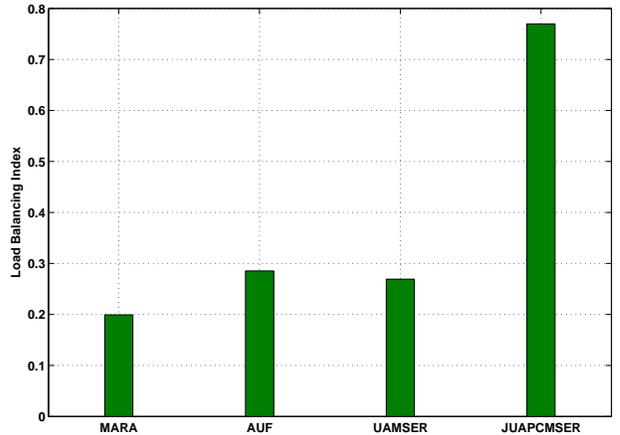}}
\caption{Load balancing levels of HCNs under different association schemes.}
\label{fig3}
\end{figure}
\par
Fig. \ref{fig3} shows the load balancing levels of HCNs under different association schemes. As shown in Fig. \ref{fig3}, compared with offloading schemes, the scheme MARA achieves a lower load balancing level since most users are attracted by BSs with high achievable rates but it is not the case for others. As revealed in Fig. \ref{fig2}, the utility function in the scheme AUF enhances the user fairness, and thus the scheme AUF achieves a high load balancing level than scheme UAMSER. In addition, the scheme JUAPCMSER achieves the highest load balancing level among all association schemes due to the impact of power control on transmit powers of BSs.
\par
Significantly, we balance network loads among different BSs to achieve a load balancing gain that the (edge) user experience is improved by balancing network loads. In order to show the gain, we mainly focus on the cumulative distribution function (CDF) of effective rates of users associated by BSs.
\begin{figure}[!t]
\centering
\centerline{\includegraphics[width=3.8in]{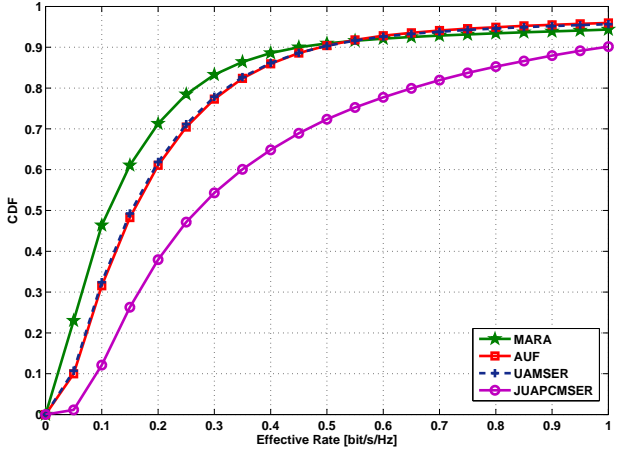}}
\caption{CDFs of effective rates of associated users under different association schemes.}
\label{fig4}
\end{figure}
\par
Fig. \ref{fig4} plots the CDFs of effective rates of associated users under different association schemes. Unlike offloading schemes, the scheme MARA lets most users select MBSs and thus results in the insufficient resources that can be utilized by users associated with these overloaded BSs. However, it is easy to know that the offloading schemes can let network resources be fully utilized and thus improve the (edge) user experience. Based on these reasons, the scheme MARA should have the most low-rate users among all association schemes. Compared with the scheme UAMSER, the scheme AUF has slightly fewer low-rate users by enhancing user fairness. As illustrated in Fig. \ref{fig4}, the scheme JUAPCMSER has the lowest low-rate users among all association schemes since the power control in this scheme mitigates interference.
\par
Considering the offloading schemes mainly improve the experiences of edge users associated with MBSs, we also investigate the CDF of effective rates of users associated with macrotier to highlight load balancing gain in an obvious manner.
\begin{figure}[!t]
\centering
\centerline{\includegraphics[width=3.8in]{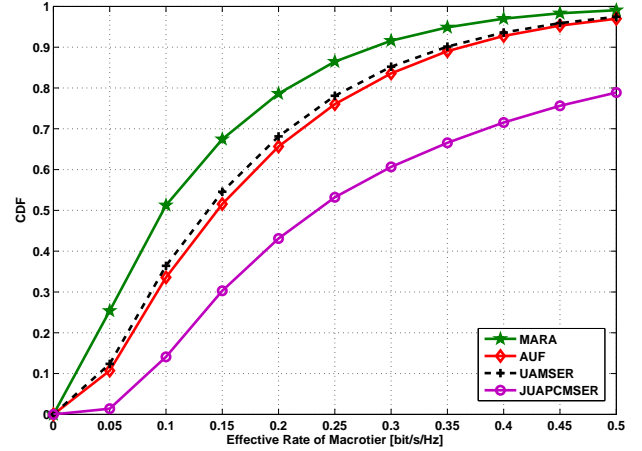}}
\caption{CDFs of effective rates of users associated with macrotier under different association schemes.}
\label{fig5}
\end{figure}
\par
Fig. \ref{fig5} plots the CDFs of effective rates of users associated with macrotier under different association schemes. As illustrated in Fig. \ref{fig5}, the scheme MARA has the most low-rate users among all association schemes, but the scheme JUAPCMSER has an oppositive result. In addition, the scheme AUF has fewer low-rate users than scheme UAMSER. Evidently, these trends are in accord with the ones illustrated in \ref{fig4}. However, we can easily find that the performance gaps among different association schemes are widen in this figure, which means the offloading scheme can more greatly improve the experiences of edge users associated with macrotier.
\par
\begin{figure}[!t]
\centering
\centerline{\includegraphics[width=3.8in]{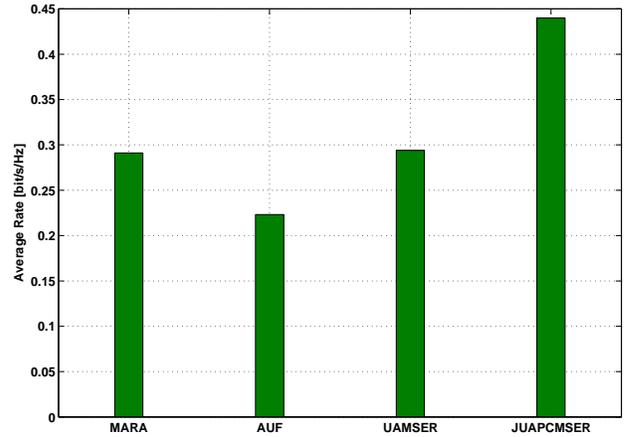}}
\caption{The average rates of different association schemes.}
\label{fig6}
\end{figure}

Fig. \ref{fig6} shows the average rates for different association schemes. Note that the average rate is denoted as the average of effective rates of all associated users. Since the schemes UAMSER and JUAPCMSER maximize the sum of effective rates, thus they achieve more higher average rates than other schemes. Moreover, the scheme JUAPCMSER has a overwhelming superiority over the scheme UAMSER since the power control in the former mitigate the interference. Although the scheme MARA has a extremely imbalanced load distribution, it has a more higher average rate than the scheme AUF. The reason for this is that the scheme MARA also has some users with very high rates even if it has the most low-rate users among all schemes. Note that these high-rates users are often associated with some underloaded BSs and thus have very more resources to be utilized.
\par
To highlight the power control gain, we also investigate the energy efficiency for different association schemes, which is denoted as the ratio of effective rate of each user to the power consumption of associated BS. Specially, the energy efficiency \cite{32} of user $k$ associated with BS $n$ is given by
\begin{equation}\label{eq32}
{{E}_{nk}}=\frac{{{R}_{nk}}}{\kappa {{p}_{n}}+{{p}^{c}}},
\end{equation}
where $\kappa$ represents the coefficient of power amplifier of BS ${n}$; ${{p}^{c}}$ is the is the circuit power consumption of BS $n$.
\par
\begin{figure}[!t]
\centering
\centerline{\includegraphics[width=3.8in]{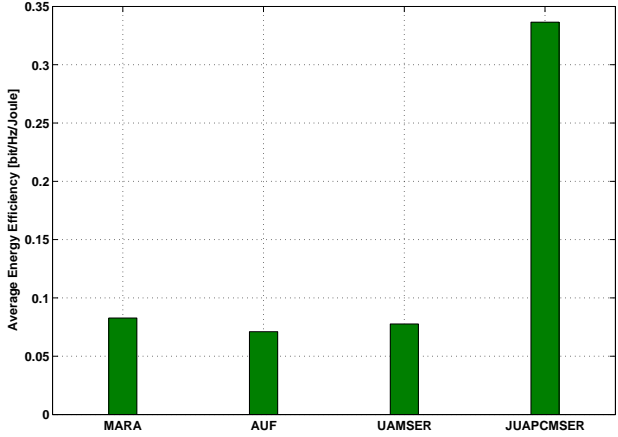}}
\caption{The average energy efficiencies of different association schemes.}
\label{fig7}
\end{figure}

Fig. \ref{fig7} investigates the average energy efficiencies for different association schemes. It is noteworthy that the average energy efficiency is denoted as the average of energy efficiencies of all associated users. Seen from Fig. \ref{fig7}, we can easily find that the scheme JUAPCMSER achieves the highest energy efficiency among all association schemes. That is because the scheme JUAPCMSER greatly improves system throughput and reduce the power consumption through power control. Compared with the scheme AUF, the scheme UAMSER has a higher energy efficiency because of higher throughput (average rate). Although the scheme MARA has a slightly lower average rate than scheme UAMSER, the fomer occupies a slightly higher energy efficiency than the latter. As mentioned in previous section, a few of users in the scheme MARA select underloaded BSs with a good channel condition, and thus may achieve very high effective rates. Maybe these rates will ensure that the scheme MARA has a higher energy efficiency than scheme UAMSER.
\par
\begin{figure}[!t]
\centering
\centerline{\includegraphics[width=3.8in]{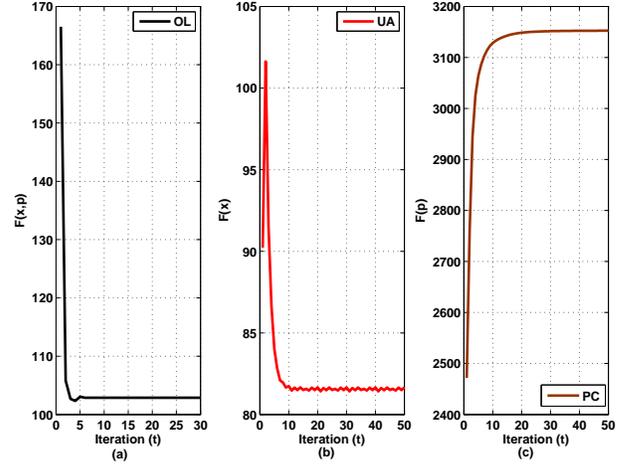}}
\caption{Convergence of proposed algorithm (Algorithm 2): (a) the convergence of outer loop (OL); (b) the convergence of user association (UA) algorithm; (b) the convergence of power control (PC) algorithm.}
\label{fig8}
\end{figure}

Fig. \ref{fig8} shows the convergence of proposed algorithm (Algorithm 2), where Fig. \ref{fig8} (a) shows the convergence of outer loop (OL); Fig. \ref{fig8} (b) shows the convergence of user association (UA) algorithm/procedure; Fig. \ref{fig8} (c) shows the convergence of power control (PC) algorithm/procedure. Since the user association algorithm and power control algorithm converges, the outer loop of Algorithm 2 will finally converge and thus the Algorithm 2 converges. As illustrated in Fig. \ref{fig8}, different iterative layers of Algorithm 2 have very high convergence rates. Evidently, the proposed algorithm can be well implemented in the practical system.
\section{Conclusion}
In this paper, we design two offloading schemes including UAMSER and JUAPCMSER, which are formulated as the problems with maximizing the sum of effective rates. Unlike scheme UAMSER, the scheme JUAPCMSER introduces power control in the association problem. Considering that the formulated problems are in mixed-integer and nonlinear forms and hard to tackle, we try to design a one-layer iterative algorithm for the scheme UAMSER, and then combine it with power control algorithm to design a two-layer iterative algorithm for the scheme JUAPCMSER. At last, we give some convergence and complexity analyses for the proposed algorithms. As shown in simulation results, the proposed schemes have superior performance than the conventional association, and the scheme JUAPCMSER achieves a higher load balancing gain and energy efficiency than conventional scheme and other offloading scheme.

\end{document}